\documentclass[conference]{IEEEtran}
\usepackage[utf8]{inputenc}
\usepackage{authblk}
\usepackage{cite}
\usepackage{authblk}
\usepackage{hyperref}
\usepackage{url}
\usepackage{breakurl}

\begin{document}
\author[1]{Clark Barrett}
\author[2]{Haniel Barbosa}
\author[3]{Martin Brain}
\author[1]{Duligur Ibeling}
\author[4]{Tim King}
\author[2]{Paul Meng}
\author[1]{\\Aina Niemetz}
\author[1]{Andres N\"{o}tzli}
\author[1]{Mathias Preiner}
\author[2]{Andrew Reynolds}
\author[2]{Cesare Tinelli}
\affil[1]{Stanford University}
\affil[2]{The University of Iowa}
\affil[3]{University of Oxford}
\affil[4]{Google}

\title{CVC4 at the SMT Competition 2018}

\maketitle

\noindent
\begin{abstract}
  This paper is a description of the CVC4 SMT solver
  as entered into the 2018 SMT Competition.
  We only list important differences from the 2017 SMT Competition version of CVC4.
  For further and more detailed information about CVC4, please
  refer to the original paper~\cite{cvc4},
  the CVC4 website~\cite{cvc4-website}, or
  the source code on
  GitHub~\cite{cvc4-github}.
\end{abstract}

\section*{Overview}

CVC4 is an efficient open-source automatic theorem prover for SMT problems. It
can be used to prove the validity (or, dually, the satisfiability) of
first-order formulas in a large number of built-in logical theories and
combinations thereof.

CVC4 is intended to be an open and extensible SMT engine, and it can be used as
a stand-alone tool or as a library, with essentially no limit on its use for
research or commercial purposes (see the section on its license below for more information).

\section*{New Features \slash\hskip .25em Improvements}

The CVC4 configuration entered in the SMT Competition 2018
is an improved and extended version of the version that entered SMT-COMP 2017.
Most notably, it features the following extensions.

\vspace{0.5ex}
\paragraph*{Floating-Point Solver}
CVC4 now features a floating point solver and thus, for the first time, enters
all FP logics of all tracks of the competition.  Its FP engine uses
SymFPU~\cite{symfpu-github} to translate floating-point operations into
bit-vector operations, which are then handed to CVC4's lazy bit-vector
engine~\cite{HBJ+14}.

\vspace{0.5ex}
\paragraph*{Eager Bit-Blasting Solver} Last year, we used
CryptoMiniSat\cite{DBLP:conf/sat/SoosNC09,cms-github} version~4 as the
back-end SAT solver for CVC4's eager bit-blasting engine.
This year, for the first time,
CaDiCaL~\cite{Biere-SAT-Competition-2017-solvers,cadical-github}
(commit id b44ce4f)
serves
as our SAT back-end for eager bit-blasting.

\vspace{0.5ex}
\paragraph*{Heuristic Approaches for Non-Linear Arithmetic}
CVC4 uses techniques for handling non-linear real and integer arithmetic
inspired by recent work by Cimatti et al~\cite{DBLP:conf/tacas/CimattiGIRS17}.
If a QF\_NIA problem cannot easily be solved with that approach, it resorts to
turning the input into a bit-vector problem.  This year, it uses CaDiCaL as the
underlying SAT solver for this approach.

\vspace{0.5ex}
\paragraph*{Quantifier Instantiation}
For unsatisfiable problems with quantifiers, 
CVC4 primarily uses conflict-based quantifier instantiation~\cite{RTd14} 
and E-matching.
CVC4 additionally implements finite model-finding techniques~\cite{RT+13-CADE} for
satisfiable problems with quantifiers.

\vspace{0.5ex}
\paragraph*{Quantified Bit-Vectors}
In~\cite{bv-cav18,DBLP:journals/corr/abs-1804-05025},
we present a novel approach for solving quantified
bit-vectors based on computing symbolic inverses
of bit-vector operators.
This approach is now the default for quantified bit-vectors in CVC4.

\vspace{0.5ex}
\paragraph*{Strings}
This year, CVC4 is entering the non-competitive
experimental division QF\_SLIA (strings). In this division,
CVC4 uses the procedure described in~\cite{DBLP:conf/cav/LiangRTBD14}
combined with a finite model-finding approach, which searches for strings
of bounded length. For handling extended string functions like string contains, substring and replace,
CVC4 uses context-dependent simplification techniques as described in~\cite{DBLP:conf/cav/ReynoldsWBBLT17}.

\section*{Configurations}

This year's version of CVC4 is entering all divisions in the main, application,
and unsat core tracks of SMT-COMP 2018. It further enters the non-competitive
experimental division QF\_SLIA (strings). All configurations are compiled with
the optional dependencies ABC~\cite{abc-website}, CLN~\cite{cln-website},
glpk-cut-log~\cite{glpk-cut-github} (a fork of GLPK~\cite{glpk-website}),
CaDiCaL, and CryptoMiniSat version~5. The commit used for all configurations is tagged
with \texttt{smtcomp2018}~\cite{smtcomp2018-tag}. For each track, we use a
binary that was compiled with different options and the corresponding run
script uses different parameters depending on the logic used in the input. For
certain logics, we try different options sequentially. For details about the
parameters used for each logic, please refer to the run scripts.

\vspace{0.5ex}
\paragraph*{Main track (CVC4-main)}
For the main track, we configured CVC4 for optimized reading
from non-interactive inputs and without proof support.
In contrast to last year's version, we do not use
a portfolio configuration for QF\_BV since the eager bit-blasting engine
with CaDiCaL as a back end in sequential configuration is more efficient.
The run script is available at~\cite{main-runscript}.

\vspace{0.5ex}
\paragraph*{Application track (CVC4-application)}
For the application track, we configured CVC4 for optimized reading from
interactive inputs and without proof support. The run script is available
at~\cite{application-runscript}.

\vspace{0.5ex}
\paragraph*{Unsat core track (CVC4-uc)}
For the unsat core track, we configured CVC4
for optimized reading from non-interactive inputs
and with proof support (required for unsat core support). The run script is
available at~\cite{uc-runscript}.

\vspace{0.5ex}
\paragraph*{Experimental (CVC4-experimental-idl-2)}
Additionally, an experimental configuration,
which features a specialized IDL solver,
enters the QF\_IDL division of the main track.
It implements
a shortest paths algorithm
as an incremental version of the Floyd-Warshall algorithm that can update
weights as new edges are added.

\section*{Copyright}
CVC4 is copyright 2009--2018 by
its authors and contributors and their institutional affiliations.
For a full list of authors, refer to the AUTHORS file
distributed with the source code~\cite{cvc4-github}.

\section*{License}

The source code of CVC4 is open and available to students, researchers,
software companies, and everyone else to study, to modify, and to redistribute
original or modified versions; distribution is under the terms of the modified
BSD license.  Please note that CVC4 can be configured (however, by default it
is not) to link against some GPLed libraries, and therefore the use of these
builds may be restricted in non-GPL-compatible projects. For more information
about CVC4's license refer to the actual license text as distributed with its
source code~\cite{cvc4-github}.

\newpage

  \bibliography{smtcomp-2018}
  \bibliographystyle{plain}

\end{document}